\documentclass{emulateapj}

\newcommand{\nad}{{Na~{\sc{i}}~D$_{1}$}}

\slugcomment{}

\shorttitle{A study of Magnetic Bright Points in the {\nad} line}
\shortauthors{D.B. Jess et al.}

\begin{document}

\title{A study of Magnetic Bright Points in the {\nad} line}

\author{D. B. Jess, M. Mathioudakis}
\affil{Astrophysics Research Centre, School of Mathematics and Physics, Queen's University, Belfast, BT7~1NN, 
Northern Ireland, U.K.}
\email{d.jess@qub.ac.uk}

\author{D. J. Christian}
\affil{Department of Physics and Astronomy, California State University, Northridge, CA 91330, U.S.A.}

\and

\author{P. J. Crockett and F. P. Keenan}
\affil{Astrophysics Research Centre, School of Mathematics and Physics, Queen's University, Belfast, BT7~1NN, 
Northern Ireland, U.K.}

\begin{abstract}
High cadence, multiwavelength, optical observations of solar magnetic bright points, captured at disk center 
using the ROSA and IBIS imaging systems on the Dunn Solar Telescope, are presented. Magnetic bright points manifesting 
in the {\nad} core are found to preferentially exist in regions containing strong downflows, in 
addition to co-spatial underlying photospheric magnetic field concentrations. Downdrafts within {\nad} bright 
points exhibit speeds of up to 7~kms$^{-1}$, with preferred structural symmetry in intensity, magnetic field and 
velocity profiles about the bright point center. Excess intensities 
associated with G-band and Ca~{\sc{ii}}~K observations of magnetic bright points reveal a power-law trend when 
plotted as a function of magnetic flux density. However, {\nad} observations of the same magnetic features 
indicate an intensity plateau at weak magnetic field strengths below $\approx$150~G, suggesting the presence of a 
two-component heating process; one which is primarily acoustic, the other predominantly magnetic. We suggest that 
this finding is related 
to the physical expansion of magnetic flux tubes, with weak field strengths ($\approx$50~G) expanding by 
$\sim$76\%, compared to a $\sim$44\% expansion when higher field strengths ($\approx$150~G) are present. 
These observations provide the first experimental evidence of rapid downdrafts in {\nad} magnetic bright 
points, and reveal the nature of a previously unresolved intensity plateau associated with these structures.
\end{abstract}


\section{Introduction}
\label{intro}
In solar physics, the {\nad} absorption profile has long been a difficult line to study, and 
indeed model. Line profile fluctuations often result from a combination of Doppler wavelength shifts, 
p-mode oscillations and other small-scale 
dynamical phenomena \citep[][]{Edm83}. Furthermore, temperature, microturbulence and magnetic field strength can affect 
the associated {\nad} line depths \citep[][]{Ath69,Sla72}. 
The formation height of the {\nad} core has often been the most difficult to interpret. \cite{Aim76} used slitless 
spectrograms to estimate a core formation height of 1300--1700~km. This is above the formation height of the 
Ca~{\sc{ii}}~K core \citep[$\approx$1200~km;][]{Bee69}, and overlaps with that of the H$\alpha$ core 
\citep[$\approx$1500~km;][]{Ver81}. However, observations indicate that {\nad} core images have little in common with 
their Ca~{\sc{ii}}~K or H$\alpha$ counterparts. \cite{Eib01} have utilized theoretical {\nad} response functions 
to suggest that the formation height varies depending on the local temperature and/or velocity. The authors estimate a 
{\nad} core formation height of $<$1000~km, based upon the non-local thermodynamic equilibrium radiative transfer 
code of \cite{Car86}. This is consistent with the work of \cite{Tom60}, \cite{Fin04} and \cite{Sim08}, who suggest 
{\nad} line-core emission originates from the upper photospheric layer ($<$800~km).

Recent {\nad} observations and simulations have revealed that {\nad} 
core brightenings sample magnetic field concentrations in the quiet solar network \citep[][]{Lee10}. 
Through a comparison of simultaneous {\sl{MDI}} magnetograms with ground-based {Na~{\sc{i}}~D$_{2}$} observations, 
\cite{Cau00} were able to suggest that sodium bright points are coincident in position, size and shape with corresponding 
magnetic patches. However, the relatively poor spatial resolution of {\sl{MDI}} ($\approx$1.2$''$) limited 
the study to network bright points in excess of $\approx$3$''$ photospheric diameter. 
Thus, it is imperative to study these types of bright features 
with an imaging system capable of 
high spatial resolution. 
In this paper, we utilize a high-cadence multiwavelength data set to search for correlations linking 
small-scale {\nad} intensity enhancements to the underlying photospheric magnetic field.

\begin{table*}
\begin{center}
\caption{ROSA/IBIS filter and cadence overview. \label{table1}}
\begin{tabular}{lccc}
~&~&~&~ \\
Filter Used			& Exposure Time & Frames per 	& Reconstructed \\
				& (ms)		& Second	& Cadence (s) \\
~&~&~&~ \\
\tableline
G-band 				& ~15 	 	& 30.3 		& 0.528        \\
Blue continuum (4170~\AA) 	& ~10 		& 30.3 		& 0.528        \\
Ca~{\sc{ii}}~K core			& 200 		& ~3.8 		& 4.224        \\
H$\alpha$ core (Zeiss)		& 240 		& ~3.8 		& 4.224        \\
LCP$^{a}$ (6302.375~\AA)	& 240 		& ~3.8 		& 4.224        \\
RCP$^{b}$ (6302.375~\AA) 	& 240 		& ~3.8 		& 4.224        \\
{\nad} (5895.94~\AA)		& ~70		& ~~2.3$^{c}$	& 39.703$^{d}$ \\
~&~&~&~ \\
\end{tabular}
\footnotesize \\
$^{a}$: Left-hand Circularly Polarized light \\
$^{b}$: Right-hand Circularly Polarized light \\
$^{c}$: Average frames per second including Fabry-Perot tuning time \\
$^{d}$: Cadence of a complete {\nad} profile scan \\
\end{center}
\end{table*}

\section{Observations}
\label{obs}
The data presented here are part of an observing sequence obtained during 13:46 -- 14:47~UT on 2009 May 28, 
with the Dunn Solar Telescope at Sacramento Peak, New Mexico. The 
Rapid Oscillations in the Solar Atmosphere \citep[ROSA;][]{Jes10b} six-camera system was employed in 
conjunction with the Interferometric BIdimensional Spectrometer \citep[IBIS;][]{Cav06} to image a 
$69.3''$~$\times$~$69.1''$ region positioned at solar disk center.   
IBIS sampled the {\nad} absorption line at 5895.94~{\AA}, incorporating nine wavelength steps with 
ten exposures per step to assist image reconstruction. 
More details of our instrumentation setup are given in table~\ref{table1}. A spatial sampling of $0.069''$ per pixel was 
used for the ROSA cameras, to match the telescope's diffraction-limited resolution in the blue continuum to that of 
the CCD. This results in images obtained at longer 
wavelengths being slightly oversampled. However, this was deemed desirable to keep the dimensions of the 
field of view the same for all ROSA cameras. IBIS employed a spatial sampling of $0.083''$ per pixel, which allowed 
ROSA's near square field-of-view to be contained within the circular aperture provided by IBIS.

During the observations, high-order adaptive optics \citep[][]{Rim04} 
were used to correct wavefront deformations in real-time. The acquired images were 
further improved through speckle reconstruction algorithms \citep[][]{Wei83, Wog08}, 
utilizing $16 \rightarrow 1$ and $10 \rightarrow 1$ restorations 
for ROSA and IBIS images, respectively. Post-reconstruction 
cadences for ROSA and IBIS are displayed in the fourth 
column of table~\ref{table1}. A full image-reconstructed IBIS scan through the {\nad} 
absorption line resulted in a cadence of 39.703~s, and includes a blueshift 
correction required due to the use of classical etalon mountings \citep[][]{Cau08}. 
To insure accurate coalignment in all 
bandpasses, broadband time series were Fourier co-registered and de-stretched using a 
$40 \times 40$ grid, equating to a $\approx 1.7''$ separation between spatial samples 
\citep[][]{Jes07,Jes08}. Narrowband images, including those from IBIS, were corrected 
by applying destretching vectors established from simultaneous broadband reference 
images \citep[][]{Rea08}. Line-of-sight magnetograms were generated 
as a difference image, normalized to their sum, of left- and right-hand circularly polarized light, obtained 
125~m{\AA} into the blue wing of the magnetically-sensitive Fe~{\sc{i}} absorption line 
at 6302.5~\AA. Utilizing high-resolution {\sl{MDI}} magnetograms of a small active region, obtained in 
conjunction with ROSA using an identical optical setup, and following the previous work of 
\cite{Ish07}, these normalized difference images were subsequently scaled to 
provide calibrated line-of-sight magnetograms. Due to the ROSA cameras 
remaining within their linear regime, and {\sl{MDI}} observations residing below their saturation 
threshold, only linear corrections were required to produce calibrated line-of-sight magnetograms 
\citep[][]{Jes10c}.

\begin{figure*}
\epsscale{1.0}
\plotone{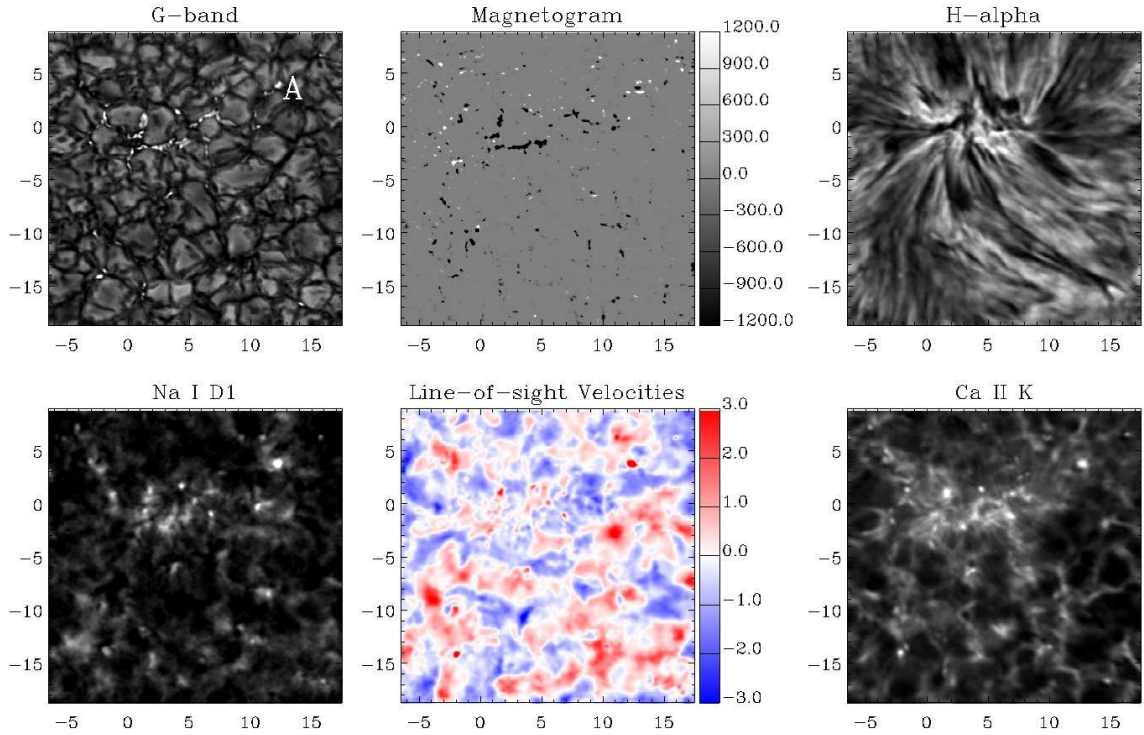}
\caption{Images from ROSA/IBIS obtained simultaneously at 14:06:21~UT. Color scales for 
the line-of-sight magnetogram and velocity maps indicate magnetic field strength (in Gauss) 
and {\nad} core velocity (in kms$^{-1}$), respectively. Artificial downdraft saturation is displayed 
in the velocity map to aid identification of MBP features. Axes are solar heliocentric 
coordinates in arcseconds. Properties related to the MBP labelled `A' are subsequently displayed 
in Figures~\ref{f2} \& \ref{f3}. 
\label{f1}}
\end{figure*}

\section{Analysis and Discussion}
\label{analy}
No active regions were present on disk during our observations, and the ROSA/IBIS field-of-view contained 
no large-scale magnetic activity. However, a small collection of magnetic bright points 
(MBPs) were visible at heliocentric co-ordinates ($3.9''$,$-1.5''$), providing an 
opportunity to examine these kiloGauss structures without the line-of-sight effects associated with near-limb observations. 
Figure~\ref{f1} shows snapshots of all the ROSA/IBIS sub-images obtained at 14:06:21~UT, and reveals 
a collection of intensity enhancements, visible in the core of the {\nad} line (lower-middle 
panel of Fig.~\ref{f1}). Ca~{\sc{ii}}~K observations were acquired through a 
1~{\AA} wide filter, with resulting emission originating from 
the upper photosphere to the lower chromosphere (see e.g. lower-right panel of Fig.~\ref{f1}). 
The {\nad} core image is an intensity map, created by establishing the 
line-profile minimum per pixel. By displaying Doppler-compensated line-center intensities, 
rather than rest-wavelength line-center intensities, brightness variations throughout the image 
are more indicative of the source function than of the velocities present in the line-forming region 
\citep[]{Lee10}. Examination of G-band (upper-left panel of Fig.~\ref{f1}) and {\nad} core 
images reveal, crucially, that 
not all {\nad} core brightenings can be directly linked to underlying MBPs. Thus, there 
may be two possible interpretations which promote efficient {\nad} core brightening. The first implies independent 
heating processes, whereby magnetic and non-magnetic brightenings are produced via unrelated mechanisms. Contrarily, another interpretation 
revolves around the same heating mechanism, but that the process is somehow more effective in highly magnetic areas, such as those 
found in MBP locations. Therefore, it is important to investigate small-scale MBPs, and establish which process 
allows enhancement of the {\nad} line above these $<2''$ G-band structures.

\begin{figure*}
\epsscale{1.0}
\plotone{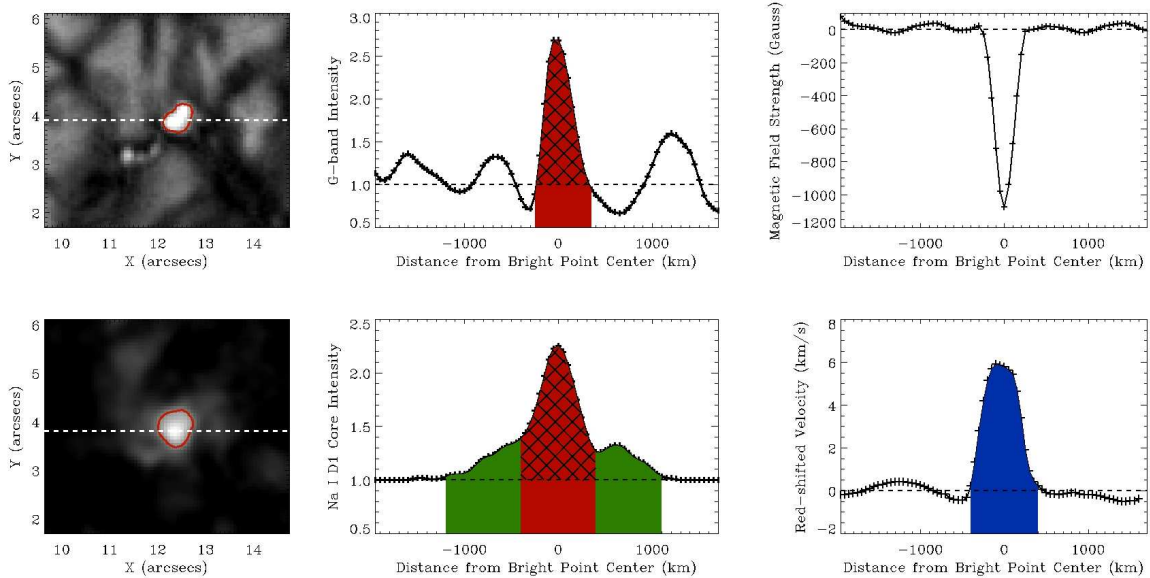}
\caption{The left-hand panels show the G-band (upper) and {\nad} core (lower) intensity images of the MBP labelled `A' 
in Figure~\ref{f1}. Overplotted in red are the pixels used to determine the excess intensity plotted in 
Figure~\ref{f4}. The remaining panels plot the one-dimensional variations in normalized G-band intensity 
(upper-middle), normalized {\nad} core intensity (lower-middle), magnetic field strength (upper-right), 
and {\nad} core downdraft velocities (lower-right), plotted as a function of distance from the center of the MBP 
along the cut indicated by the horizontal dashed lines in the left-hand panels. 
Green shading in the normalized {\nad} core intensity plot indicate the extent of the halo around the MBP, 
while red shading marks the locations of the central MBP. Blue shading reveals regions of 
non-zero downdraft velocities, synonymous with {\nad} core MBP enhancements. Cross-hatched regions display the 
values used when determining the excess intensity.
\label{f2}}
\end{figure*}

Information contained within the {\nad} core velocity maps, generated through Doppler-shifts of the profile minimum, begins to unlock this mystery. 
An inspection of Figure~\ref{f1} reveals a preference 
for {\nad} and G-band MBP structures to exist in downflow regions (up to 7~kms$^{-1}$). Furthermore, Ca~{\sc{ii}}~K core and H$\alpha$ 
core intensity enhancements are often observed to co-exist in locations of strong downflows ($>$3~kms$^{-1}$) and 
longitudinal magnetic fields. 
Figure~\ref{f2} shows 
a slice through a typical high red shift feature, displaying its G-band intensity, {\nad} core 
intensity, magnetic field strength, and {\nad} line-core velocity as a function of distance across the 
underlying MBP.

\begin{figure*}
\epsscale{1.0}
\plotone{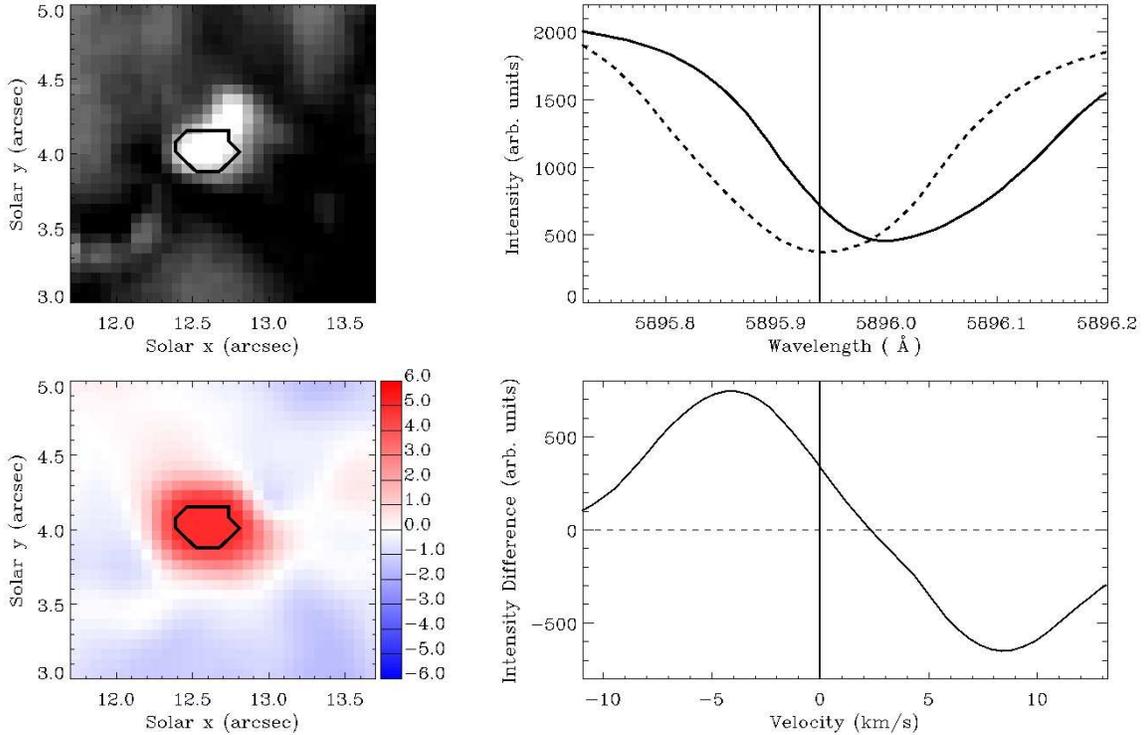}
\caption{{\it{Top left:}} G-band intensity image of the MBP labelled `A' in Figure~\ref{f1}. {\it{Lower left:}} Line-of-sight 
velocity map of the same MBP, obtained from Doppler wavelength shifts of the {\nad} profile. The color scale indicates 
flow velocities in kms$^{-1}$, where red and blue colors represent downflows and upflows, respectively. In each panel, the solid black 
contour indicates regions where downflow velocities exceed 4~kms$^{-1}$. 
{\it{Top right:}} Averaged {\nad} profile (solid line) generated from locations contained within the contours displayed in the left panels of 
this Figure, overplotted with a 
rest-wavelength {\nad} profile (dashed line). 
{\it{Lower right:}} The difference between the two profiles. In both right-hand panels, the solid vertical line represents the {\nad} 
rest wavelength at 5895.94~{\AA}.
\label{f3}}
\end{figure*}

In the top-right panel of Figure~\ref{f3}, we present an averaged {\nad} line profile (solid line) from the 
central region of a MBP, labelled `A' in Figure~\ref{f1}, where high redshifts are observed. 
A dashed line represents a {\nad} profile, which has been spatially, and temporally, averaged over the entire 
field-of-view and time series (in excess of $9 \times 10^{7}$ individual profiles). It is evident that 
the MBP profile has been significantly Doppler shifted when compared to the average {\nad} profile. 
The difference between these two profiles is displayed in the bottom-right panel of Figure~\ref{f3}, and reveals a blue bump at 
$\approx$$-$4~kms$^{-1}$, alongside a red shifted dip at $\approx$8~kms$^{-1}$. These asymmetries 
are indicative of shock formation below the chromospheric canopy, and are in remarkable agreement 
with the simulations of \cite{Lee10}.

\cite{Rou06} investigated velocity flows associated with the {\nad} line in strong magnetic field 
regions surrounding active regions and sunspots. These authors detected an abundance of downflows 
in these locations, with velocities nearing 2~kms$^{-1}$. As shown in the lower-right 
panel of Figure~\ref{f2}, the {\nad} core velocities presented here are considerably higher, 
reaching $\approx$7~kms$^{-1}$ at the center of the MBP. 
Even artificially 
binning our data to mimic the resolution of \cite{Rou06} would not reduce the peak 
{\nad} Doppler velocities to below $3$~kms$^{-1}$. However, the spectral resolution of IBIS 
(23~m{\AA} FWHM) is considerably better than the resolution obtained by the 
Multichannel Subtractive Double Pass system (144~m{\AA} FWHM) employed by \cite{Rou06}, and 
allows a higher degree of precision when determining line-core velocities. Quantitatively, 
the Doppler velocity error, $err$, can be defined as $err = \pm \frac{{\mathrm{FWHM}}~c}{2 \lambda}$ 
\citep[][]{Lan08}, where $\lambda$ is the line-core wavelength and $c$ is the speed of light. This 
produces an associated IBIS velocity error of $\pm$0.5~kms$^{-1}$, compared to an error of 
$\pm$3.7~kms$^{-1}$ in the results of \cite{Rou06}. In addition, {\nad} downdraft velocities 
approaching 7~kms$^{-1}$ are consistent with the magneto-hydrodynamic simulations of \cite{Lee10}. 
Thus, 
for the first time, the combined high spatial and spectral resolutions of speckle-reconstructed 
{\nad} IBIS data sets allow fine-scale Doppler velocities 
to be evaluated with an unprecedented degree of precision.

From the lower panels of Figure~\ref{f2}, we see that MBP structures found in {\nad} core 
imaging demonstrate an increased width over simultaneous G-band observations. This is 
consistent with the ``aureoles'', or halos, found by \cite{Lee10}, who interpret 
these additional brightenings as an artifact of strong radiative scattering. It must be noted 
that the decrease in normalized intensity to values below 1.0, at the edges of the G-band 
MBP, is due to intensity sampling of the dark inter-granular lane, where the MBP resides. 
Examination of Figure~\ref{f2} suggests that the fastest downdrafts are co-spatial with the center 
of the MBP, and that the velocities drop to approximately zero at the edge of the structure. Furthermore, 
the strongest longitudinal magnetic field strength is also co-spatial with the center of the 
MBP, and correlates well in size with the width of the G-band structure. 
Following analysis of all MBPs observed with red shifts exceeding 3~kms$^{-1}$ and underlying magnetic 
field concentrations, prevailing symmetry is found in intensity, velocity and magnetic-field strength, as a function 
of radial distance from the center of the MBP. This indicates the preference 
for an ordered structure, whereby maximum values are attributed to the center of the MBP, and steeply 
decaying measurements are found at increasing distances from the center.

\begin{figure}
\epsscale{1.0}
\plotone{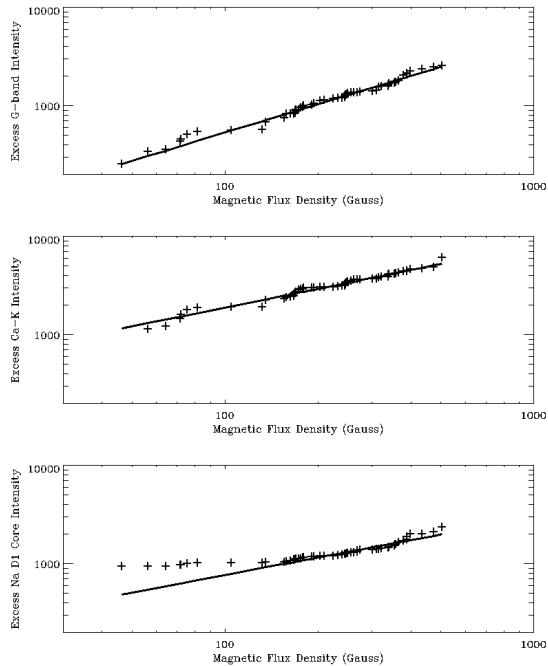}
\caption{Intensity excess for each MBP containing downdraft velocities exceeding 3~kms$^{-1}$, plotted as a function of 
MBP-averaged magnetic flux density for G-band (top), Ca~{\sc{ii}}~K (middle) and {\nad} (bottom) observations, and 
displayed on logarithmic axes. In each panel, power-law fits have been applied to the data which best minimizes the 
chi-squared error statistic. 
\label{f4}}
\end{figure}

Investigating the intensity excess of MBPs, which correspond to all MBPs containing strong ($>$3~kms$^{-1}$) downflows 
and G-band inter-granular brightenings, reveals a 
correlation with the underlying magnetic field concentrations. 
Figure~\ref{f4} reveals how the intensity excess of G-band, Ca~{\sc{ii}}~K and {\nad} observations 
varies as a function of the magnetic flux density, $\Phi$, defined as the total magnetic field strength divided by the number of 
occupied pixels. The excess intensity, $\Delta I$, is determined as an averaged pixel intensity for the entire MBP, following 
a subtraction of intensities equal to the modal value observed in quiet areas (central panels of Figure~\ref{f2}). For G-band images, 
MBPs are isolated from granules and inter-granular lanes using the automated routines of \cite{Cro09}. These isolated pixels 
also provide the structures used to determine the magnetic flux density from the line-of-sight magnetograms. {\nad} core MBP 
observations consist of a bright central structure, surrounded by a dimmer halo, resulting from radiative scattering \citep[]{Lee10}. 
The central MBP is isolated from the halo by utilizing simultaneous {\nad} core velocity maps. The lower-right panel of Figure~\ref{f2} reveals how 
the downdraft velocities drop to approximately zero at the edge of the bright central MBP. Therefore, pixels where the downdraft velocity 
is non-zero are used to isolate the central MBP regions, thus minimizing contributions to the excess intensity caused by radiative 
scattering, which is only weakly dependent on the magnetic field strength.

In the case of G-band observations (upper panel 
of Fig.~\ref{f4}), a linear dependence observed in the log--log plot is in agreement with \cite{Bec07}, who detect identical 
trends for G-band brightenings in active region moats. 
Through analyses of network bright points ($>$3$''$ in diameter) and large-scale active region structures 
($>$15$''$ in diameter), \cite{Cau00} and \cite{Sch89} were able to 
demonstrate a similar trend for these features when observed in the {Na~{\sc{i}}~D$_{2}$} and Ca~{\sc{ii}}~K lines, respectively. 
Indeed, for the magnetic flux densities presented here (above $\approx$150~G in the case of sodium), the calcium, G-band and sodium excesses 
(Fig.~\ref{f4}) are best fitted by a power law of the type 
$\Delta I \propto \Phi^{\beta}$, where $\beta$ is the power-law exponent. 
Here, we determine $\beta = 0.61 \pm 0.02$ for Ca~{\sc{ii}}~K observations, and $\beta = 0.59 \pm 0.03$ in the case of {\nad} data. 
These values of power-law exponent are in agreement with the previous {Na~{\sc{i}}~D$_{2}$} study of 
\cite{Cau00}, and remain consistent with exponents found in Ca~{\sc{ii}}~K investigations \citep[][]{Sku75, Nin98}. While these consistencies with 
previously published work may seem initially superfluous, the spatial scale over which these results agree is extraordinary. \cite{Sch89} 
investigated large-scale magnetic structures contained within an active region, often with spatial sizes approaching $70''$ in diameter. 
The MBPs investigated here consist of structures as small as $\approx$0.2$''$ in diameter, more than two 
orders of magnitude smaller than those studied previously. 

At magnetic flux densities below $\approx$150~G, there is an appreciable ``turn-off'' from the power-law fit applied to the {\nad} 
observations. The lower panel of Figure~\ref{f4} reveals how the {\nad} intensity remains 
elevated, even as the magnetic flux density begins to diminish. Through analysis of Ca~{\sc{ii}}~K observations, \cite{Sch89} suggest 
large deviations in line-core intensity, at relatively 
small magnetic flux densities, may be a result of a large inclination of the magnetic flux tubes, or small-scale energetic phenomena. 
Inclination effects on the data presented here are not believed to be significant for two reasons. 
Firstly, the inclination angle of MBPs, with respect to the vertical, is typically less than $10^{\circ}$ \citep[][]{San94}. Secondly, due 
to the observations being acquired at disk center, line-of-sight effects are minimized. This allows photospheric 
structures to remain co-spatial when observed in chromospheric passbands, and since no feature migration is found between G-band and Ca~{\sc{ii}}~K 
atmospheric heights, further solidifies the presence of minimal inclination angles. 
Additionally, 
the intensities of MBPs corresponding to G-band, Ca~{\sc{ii}}~K and {\nad} observations show no temporal variations that may be suggestive 
of microflare \citep[][]{Jes10a} and/or Ellerman bomb \citep[][]{Ell17} activity. 
Through examination of photospheric G-band MBPs, we observe the physical area of such 
structures to reduce as the magnetic field becomes weaker, corroborating the observational results of \cite{Ish07}.
Contrarily, we find MBPs, which correspond to magnetic flux densities below the turn-off at $\approx$150~G, 
show similar two-dimensional areas when viewed in the {\nad} line core. 
An average {\nad} core area of 0.11$\pm$0.01~Mm$^{2}$ (47$\pm$4~pixels) is found for 
these structures, suggesting their excess intensity plateau below $\approx$150~G may 
be related to the physical expansion of magnetic flux tubes. \cite{Sol99} utilize models based upon infrared spectral lines to 
infer how small magnetic flux tubes expand more rapidly as a function of atmospheric height, when compared to larger magnetic 
structures.  Thus, in order to maintain relatively constant physical sizes 
present in {\nad} observations, it is 
necessary for the physically smaller (and weaker) magnetic field concentrations in the photosphere to expand at a greater rate. To 
quantify this result, we find MBPs at the lowest magnetic flux densities ($\approx$50~G) show a $\sim$76\% increase in area between G-band 
and {\nad} observations, compared to a $\sim$44\% increase for MBPs near the $\approx$150~G turn-off.

A discontinuity in the correlation between the magnetic field and {\nad} intensity, for magnetic flux densities below $\approx$150~G, 
suggests that these MBP intensities are independent of the underlying magnetic field strength. 
Combining {\nad} observations with 3D non-LTE simulations, \cite{Lee10} conclude that the {\nad} line is not chromospheric in nature, 
and instead samples the atmospheric layer below the magnetic canopy, where magneto-acoustic shocks are believed to dominate. 
Our findings may therefore indicate a two-component heating process; one which is primarily acoustic and dominates the heating 
of low magnetic field strength areas, while the other process is predominantly magnetic in nature. 
Within the presented data, MBPs with velocities exceeding 3~kms$^{-1}$ represent the dominant population. However, as 
MBPs can exist with velocities below this threshold, their excess intensity behaviour, as a function of magnetic flux density, may differ 
as a result. We note that two-component heating models 
have previously been proposed to interpret observed chromospheric flux densities in the most inactive solar-type stars \citep[][]{Sch89b, Math92}.


\section{Concluding Remarks}
\label{conc}

Joint observations, acquired with the Rapid Oscillations in the Solar Atmosphere (ROSA) and 
Interferometric BIdimensional Spectrometer (IBIS) imaging systems, have enabled unprecedented 
views of solar magnetic bright points (MBPs). These data, obtained simultaneously with 
high spatial, temporal and spectral resolutions, have revealed how MBPs manifesting 
in the {\nad} core are found to preferentially exist in regions containing downflows, 
in addition to co-spatial underlying magnetic field concentrations. Downdrafts found within 
{\nad} bright points exhibit speeds of up to 7~kms$^{-1}$, with preferred structural symmetry 
in intensity, magnetic field and velocity profiles about the center of the MBP. Excess intensities 
associated with G-band and Ca~{\sc{ii}}~K observations of MBPs demonstrating downdrafts exceeding 
3~kms$^{-1}$ reveal a power-law trend when 
plotted as a function of magnetic flux density. However, {\nad} observations of the same magnetic features 
indicate an intensity plateau at weak magnetic field strengths below $\approx$150~G. We interpret this phenomena 
as the physical expansion of magnetic flux tubes, with weak field strengths ($\approx$50~G) expanding by 
$\sim$76\%, compared to $\sim$44\% expansion for structures at higher magnetic field strengths ($\approx$150~G).

\acknowledgments
DBJ thanks STFC for a Post-Doctoral Fellowship. 
DJC thanks CSUN for start-up funding. 
PJC is grateful to NIDEL for a PhD studentship. 
FPK thanks AWE Aldermaston for the William Penney Fellowship. 




\begin{thebibliography}{}
\bibitem[Aimanova \& Gulyaev(1976)]{Aim76}
Aimanova, G.~K., \& Gulyaev, R.~A., 1976, Soviet Astronomy, 20, 201 
\bibitem[Athay \& Canfield(1969)]{Ath69}
Athay, R.~G., \& Canfield, R.~C., 1969, \apj, 156, 695 
\bibitem[Beebe \& Johnson(1969)]{Bee69}
Beebe, H.~A., \& Johnson, H.~R., 1969, \solphys, 10, 79
\bibitem[Beck et~al.(2007)]{Bec07}
Beck, C., Bellot Rubio, L.~R., Schlichenmaier, R., S{\"{u}}tterlin, P., 2007, \aap, 472, 607 
\bibitem[Carlsson(1986)]{Car86}
Carlsson, M., 1986, Uppsala Astronomical Observatory Reports, 33  
\bibitem[Cauzzi et~al.(2000)]{Cau00}
Cauzzi, G., Falchi, A., \& Falciani, R., 2000, \aap, 357, 1093 
\bibitem[Cauzzi et~al.(2008)]{Cau08}
Cauzzi, G., et al., 2008, \aap, 480, 515 
\bibitem[Cavallini(2006)]{Cav06}
Cavallini, F., 2006, \solphys, 236, 415 
\bibitem[Crockett et~al.(2009)]{Cro09}
Crockett, P.~J., Jess, D.~B., Mathioudakis, M., \& Keenan, F.~P., 2009, \mnras, 397, 1852 
\bibitem[Edmonds \& Hsu(1983)]{Edm83}
Edmonds, F.~N., Jr., \& Hsu, J.-C., 1983, \solphys, 83, 217 
\bibitem[Eibe et~al.(2001)]{Eib01}
Eibe, M.~T., Mein, P., Roudier, T., \& Faurobert, M., 2001, \aap, 371, 1128 
\bibitem[Ellerman(1917)]{Ell17}
Ellerman, F., 1917, \apj, 46, 298 
\bibitem[Finsterle et~al.(2004)]{Fin04}
Finsterle, W., Jefferies, S.~M., Cacciani, A., Rapex, P., \& McIntosh, S.~W., 2004, \apjl, 613, L185 
\bibitem[Ishikawa et~al.(2007)]{Ish07}
Ishikawa, R., Tsuneta, S., Kitakoshi, Y., Katsukawa, Y., et al., 2007, \aap, 472, 911 
\bibitem[Jess et~al.(2007)]{Jes07}
Jess, D.~B., McAteer, R.~T.~J., Mathioudakis, M., Keenan, F.~P., Andic, A., \& Bloomfield, D.~S., 2007, \aap, 476, 971 
\bibitem[Jess et~al.(2008)]{Jes08}
Jess, D.~B., Mathioudakis, M., Crockett, P.~J., \& Keenan, F.~P., 2008, \apjl, 688, L119 
\bibitem[Jess et~al.(2010a)]{Jes10a}
Jess, D.~B., Mathioudakis, M., Browning, P. K., Crockett, P.~J., \& Keenan, F. P., 2010a, \apjl, 712, L111 
\bibitem[Jess et~al.(2010b)]{Jes10b}
Jess, D.~B., Mathioudakis, M., Christian, D.~J., Keenan, F.~P., Ryans, R.~S.~I., \& Crockett, P.~J., 2010b, \solphys, 261, 363 
\bibitem[Jess et~al.(2010c)]{Jes10c}
Jess, D.~B., Mathioudakis, M., Christian, D.~J., Crockett, P.~J., \& Keenan, F.~P., 2010c, in prep. 
\bibitem[Langangen et~al.(2008)]{Lan08}
Langangen, {\O}., Rouppe van der Voort, L., \& Lin, Y., 2008, \apj, 673, 1201 
\bibitem[Leenaarts et~al.(2010)]{Lee10}
Leenaarts, J., Rutten, R.~J., Reardon, K., Carlsson, M., \& Hansteen, V., 2010, \apj, 709, 1362
\bibitem[Mathioudakis \& Doyle(1992)]{Math92}
Mathioudakis, M., \& Doyle, J.~G., 1992, \aap, 262, 523
\bibitem[Nindos \& Zirin(1998)]{Nin98}
Nindos, A., \& Zirin, H., 1998, \solphys, 179, 253 
\bibitem[Reardon \& Cavallini(2008)]{Rea08}
Reardon, K.~P., \& Cavallini, F., 2008, \aap, 481, 897 
\bibitem[Rimmele(2004)]{Rim04}
Rimmele, T.~R., 2004, \procspie, 5490, 34
\bibitem[Roudier et~al.(2006)]{Rou06}
Roudier, T., Malherbe, J.~M., Moity, J., Rondi, S., Mein, P., \& Coutard, C., 2006, \aap, 455, 1091 
\bibitem[S{\'{a}}nchez Almeida \& Mart{\'{i}}nez Pillet(1994)]{San94}
S{\'{a}}nchez Almeida, J., \& Mart{\'{i}}nez Pillet, V., 1994, \apj, 424, 1014 
\bibitem[Schrijver et~al.(1989a)]{Sch89}
Schrijver, C.~J., Cote, J., Zwaan, C., \& Saar, S.~H., 1989a, \apj, 337, 964 
\bibitem[Schrijver et~al.(1989b)]{Sch89b}
Schrijver, C.~J., Dobson, A.~K., \& Radick, R.~R, 1989b, \apj, 341, 1035
\bibitem[Simoniello et~al.(2008)]{Sim08}
Simoniello, R., Jim{\'{e}}nez-Reyes, S.~J., Garc{\'{\i}}a, R.~A., \& Pall{\'{e}}, P.~L., 2008, Astronomische Nachrichten, 329, 494 
\bibitem[Skumanich et~al.(1975)]{Sku75}
Skumanich, A., Smythe, C., \& Frazier, E.~N., 1975, \apj, 200, 747 
\bibitem[Slaughter \& Wilson(1972)]{Sla72}
Slaughter, C.~D., \& Wilson, A.~M., 1972, \solphys, 24, 43 
\bibitem[Solanki et~al.(1999)]{Sol99}
Solanki, S.~K., Finsterle, W., R{\"u}edi, I., \& Livingston, W., 1999, \aap, 347, L27 
\bibitem[Tomita(1960)]{Tom60}
Tomita, Y., 1960, \pasj, 12, 524 
\bibitem[Vernazza et~al.(1981)]{Ver81}
Vernazza, J.~E., Avrett, E.~H., \& Loeser, R., 1981, \apjs, 45, 635 
\bibitem[Weigelt \& Wirnitzer(1983)]{Wei83}
Weigelt, G., \& Wirnitzer, B., 1983, Optics Letters, 8, 389
\bibitem[W{\"o}ger et~al.(2008)]{Wog08}
W{\"o}ger, F., von der L{\"u}he, O., \& Reardon, K., 2008, \aap, 488, 375 
\end{thebibliography}
\end{document}